\documentclass[conference,letterpaper]{IEEEtran}
\usepackage{geometry}
\usepackage{cite}
\usepackage{amsmath,amssymb,amsfonts}
\usepackage{algorithmic}
\usepackage{graphicx}
\usepackage{textcomp}
\usepackage{xcolor}
\usepackage{url}
\usepackage{booktabs}
\usepackage{subcaption}
\usepackage{tikz}
\usepackage{tikzpagenodes}
\usetikzlibrary{%
  arrows.meta
}

\def\BibTeX{{\rm B\kern-.05em{\sc i\kern-.025em b}\kern-.08em
    T\kern-.1667em\lower.7ex\hbox{E}\kern-.125emX}}

\IEEEoverridecommandlockouts

\title{Design and Autonomous Stabilization of a Ballistically-Launched Multirotor}

 \author{\IEEEauthorblockN{Amanda Bouman$^{1}$, Paul Nadan$^{2}$, Matthew Anderson$^{3}$, Daniel Pastor$^{1}$, \\ Jacob Izraelevitz$^{3}$, Joel Burdick$^{1}$, and Brett Kennedy$^{3}$\vspace{-0.7cm}}

\thanks{$^{1}$Amanda Bouman, Daniel Pastor and Joel Burdick are with the Department of Mechanical and Civil Engineering,
        California Institute of Technology, Pasadena, CA 91125, USA,
        {\tt\small abouman@caltech.edu, dpastorm@caltech.edu, jwb@robotics.caltech.edu}}%
\thanks{$^{2}$Paul Nadan is with Olin College of Engineering, Needham, MA 02492, 
        {\tt\small pnadan@olin.edu}}%
\thanks{$^{3}$Matthew Anderson, Jacob Izraelevitz and Brett Kennedy are with the Jet Propulsion Laboratory, California Institute of Technology, Pasadena, CA 91109 
        {\tt\small jacob.izraelevitz@jpl.nasa.gov, bkennedy@jpl.nasa.gov}}%
\thanks{DISTRIBUTION STATEMENT A (Approved for Public Release, Distribution Unlimited)}%
}

\begin{document}

\maketitle
\thispagestyle{empty}
\pagestyle{empty}

\begin{abstract}

Aircraft that can launch ballistically and convert to autonomous, free-flying drones have applications in many areas such as emergency response, defense, and space exploration, where they can gather critical situational data using onboard sensors. This paper presents a ballistically-launched, autonomously-stabilizing multirotor prototype (\emph{SQUID} - \emph{Streamlined Quick Unfolding Investigation Drone}) with an onboard sensor suite, autonomy pipeline, and passive aerodynamic stability. We demonstrate autonomous transition from passive to vision-based, active stabilization, confirming the multirotor's ability to autonomously stabilize after a ballistic launch in a GPS-denied environment.

\end{abstract}


\section{INTRODUCTION}
Unmanned fixed-wing and multirotor aircraft are usually launched manually by an attentive human operator. Aerial systems that can instead be launched ballistically without operator intervention will play an important role in emergency response, defense, and space exploration where situational awareness is often required, but the ability to conventionally launch aircraft to gather this information is not available. 

Firefighters responding to massive and fast-moving fires could benefit from the ability to quickly launch drones through the forest canopy from a moving vehicle. This eye-in-the-sky could provide valuable information on the status of burning structures, fire fronts, and safe paths for rapid retreat. Likewise, military personal in active engagements could quickly deploy aerial assets to gather information as the situation evolves. 

\begin{figure}[t]
	\centering
	\includegraphics[angle=0,width=0.3\linewidth, trim=1000 100 1000 200,clip]{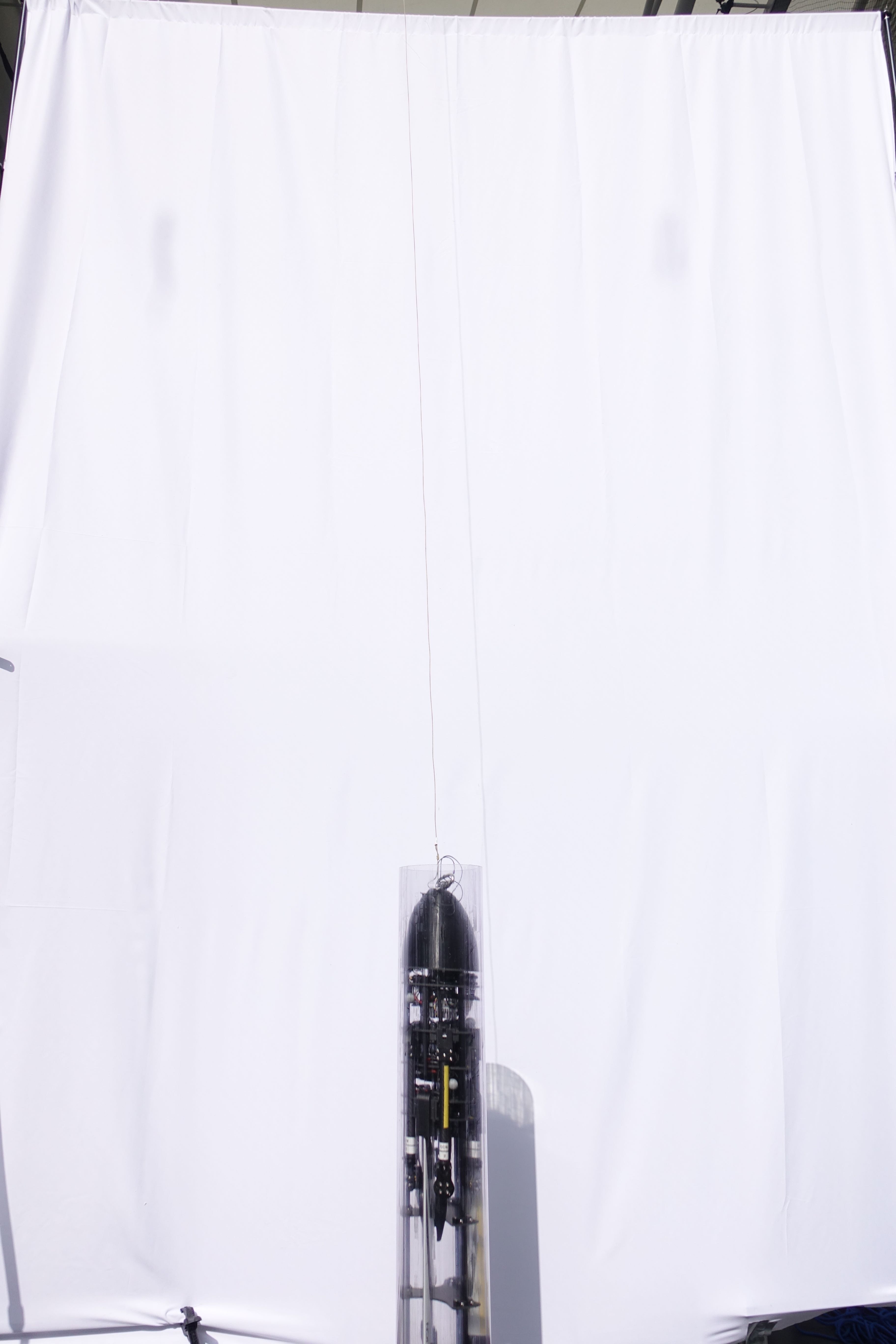}
	\includegraphics[angle=0,width=0.3\linewidth, trim=1000 100 1000 200,clip]{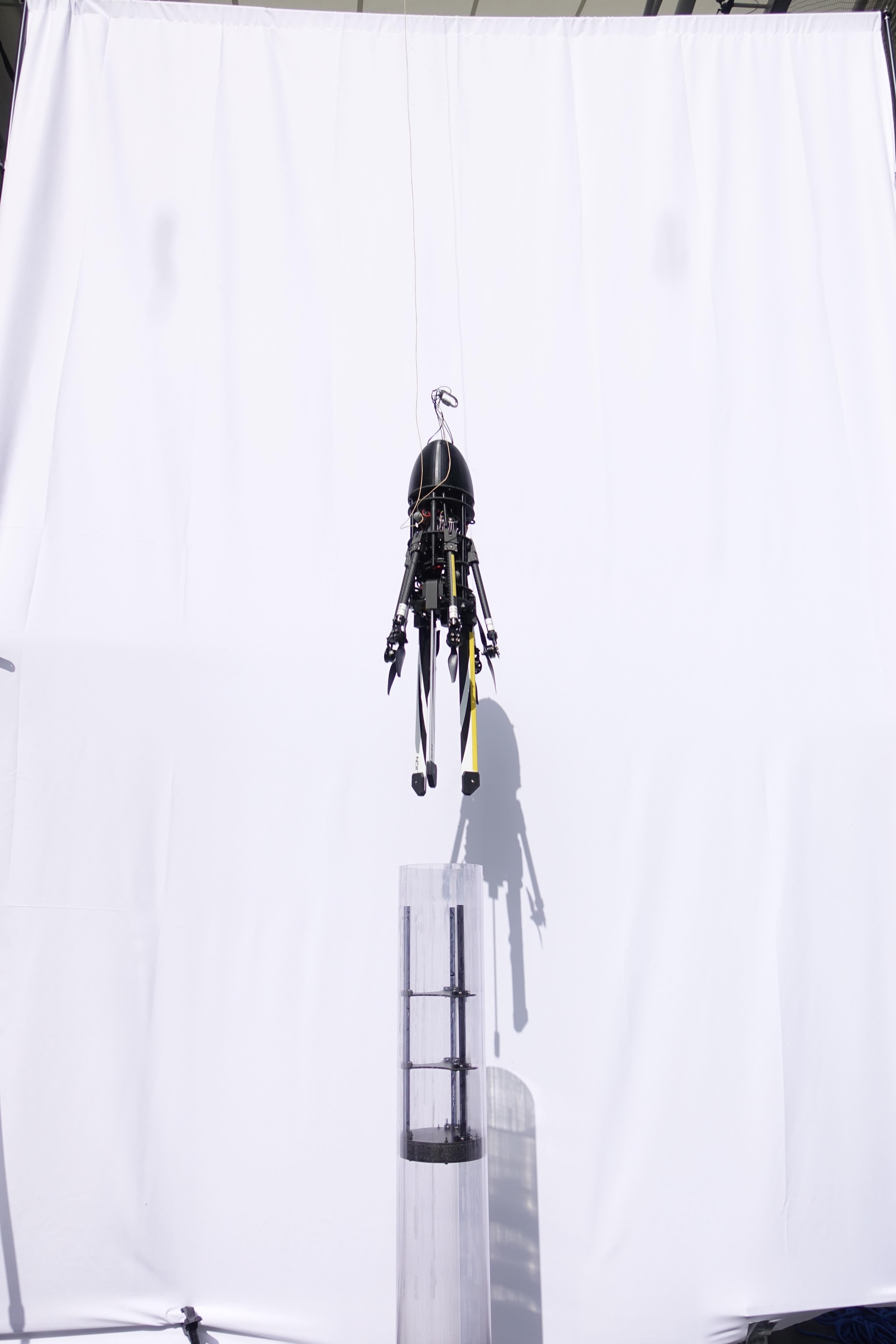}
	\includegraphics[angle=0,width=0.3\linewidth, trim=1000 100 1000 200,clip]{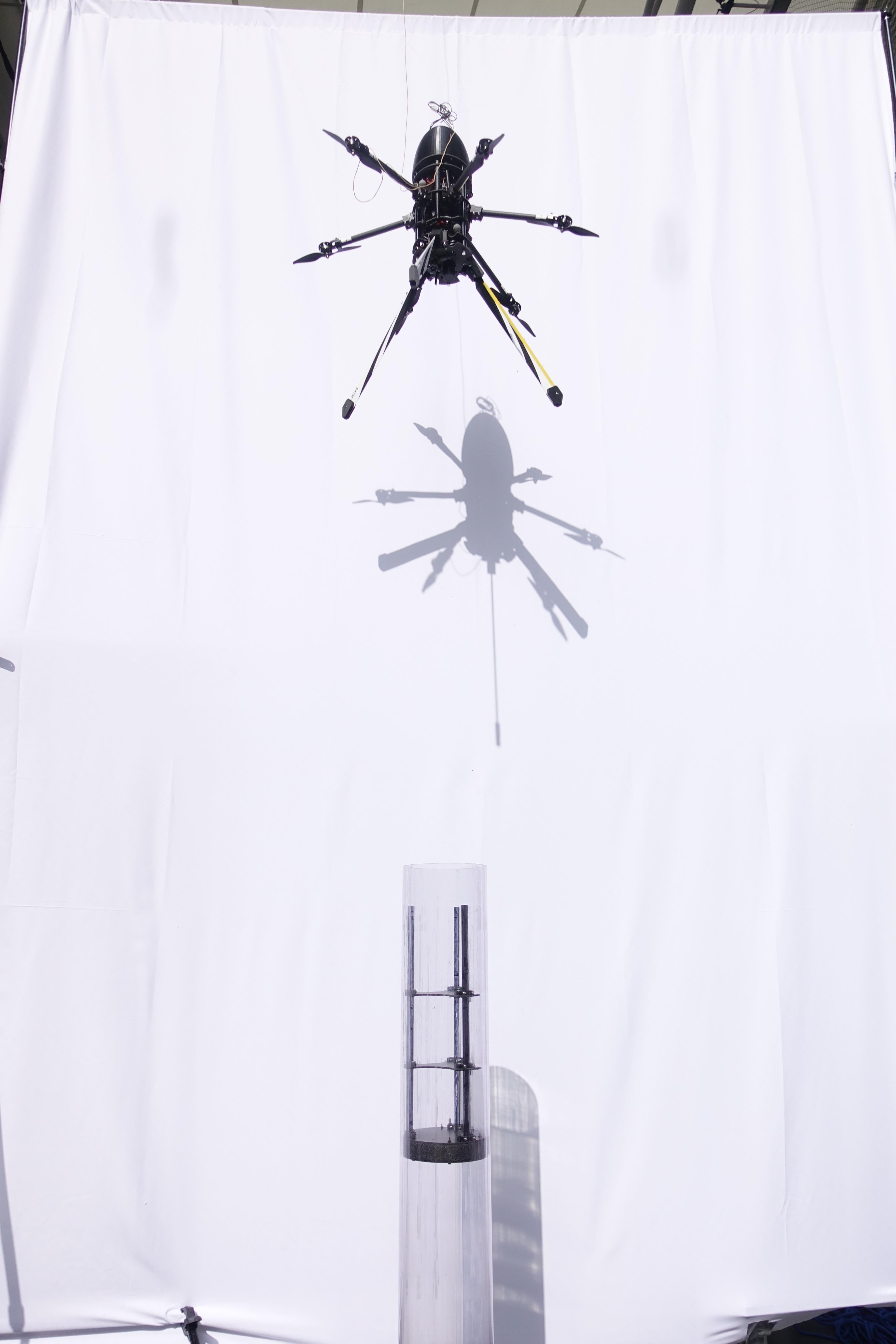}
	\caption{Launching \emph{SQUID}: inside the launcher tube (left), deploying the arms and fins (center), and fully-deployed configuration (right). Note the slack in the development safety tether and how the carriage assembly remains in the tube throughout launch. Each picture is 82 ms apart.}
	\label{fig:fernando_mother_cake}
	\vspace{-0.3cm}
\end{figure}

Ballistic launches also provide unique opportunities in the exploration of other bodies in the solar system. The \emph{Mars Helicopter Scout} (MHS), which is to be deployed from the  Mars 2020 rover, will provide the first powered flight on another solar system body in history~\cite{bib:mars}. MHS greatly expands the data collection range of the rover, however, it has a multistep launch sequence that requires flat terrain. The addition of a ballistic, deterministic launch system for future rovers or landers would physically isolate small rotorcraft from the primary mission asset.  Aerial launch technology would even enable the aircraft to deploy directly from the entry vehicle during decent and landing, enabling it to land and explore sites a great distance from the rover.  

Multirotor aircraft are advantageous over fixed-wing systems as they can hover in place and aggressively maneuver in cluttered environments to achieve greater vantage points.  However, the rotating blades of the multirotor are  a hazard to nearby personnel (who may be distracted by other obligations), a problem which is particularly present if the system is to launch autonomously without human supervision. In these situations, multirotor aircraft operating in crowded and rapidly changing environments need a precise, highly deterministic, and fully autonomous takeoff method to achieve a safe operating altitude away from assets and personnel.

In the application scenarios described above, ideally the multirotor is stored for extended periods of time ("containerized") before being launched quickly, safely, and autonomously. Furthermore, when deployed from a moving vehicle, the drone must be aerodynamically stable to avoid tumbling when exposed to sudden crosswinds. Most current drone designs however are slow to deploy, require user intervention prior to takeoff, and cannot be deployed from fast-moving vehicles. Current foldable designs also require the user to manually unfold the arms which slows the process and puts the user at risk if the multirotor prematurely activates. A multirotor that can launch from a simple tube and autonomously transition to flight would solve many of the shortcomings of conventional drone deployment strategies. 

While mature tube-launched fixed-wing aircraft are already in active use~\cite{raytheonCoyote:online,uvisionHero:online,leonardo:online}, tube-launched rotorcraft (both co-axial and multirotor) are much rarer and primarily still in development. Several consumer drones (e.g., the DJI \textit{Mavic} series~\cite{Mavic2th36:online} and Parrot \textit{Anafi}  ~\cite{DroneCam11:online}) can be folded to occupy a small volume, but these designs cannot fit smoothly inside a launch system, and the unfolding is manual. Other manually unfolding rotorcraft can achieve a cylindrical form factor like \emph{SQUID}: the \textit{Power Egg} from Power Vision folds into an egg shape~\cite{PowerEgg6:online}, the \textit{LeveTop} drone folds into a small cylinder~\cite{LeveTopT73:online}, and the coaxially designed \textit{Sprite} from Ascent Aerosystems packs into a cylinder shape~\cite{AscentAe83:online}. Automatic in-flight unfolding mechanisms for quadrotors, using both active~\cite{scaramuzza_foldable_drone_2019} and passive~\cite{mueller_foldable_drone_2019} actuation, have been developed for the traversal of narrow spaces. However, to enable the ability to ballistically launch like \emph{SQUID}, these existing foldable platforms must be redesigned to withstand launch loads and maintain passive aerodynamic stability post-launch. Ballistically-launched aerial systems that combine an aerodynamically stable structure and a foldable airfoil system have been developed in coaxial rotorcraft~\cite{gnemmi2017conception} and multirotor~\cite{henderson2017towards} formats, but both designs are still in the theoretical design phase, and have yet to demonstrate a transition from ballistic to stabilized flight. 


In previous work~\cite{SQUID:IROS19}, we introduced a small \emph{SQUID} prototype, a folding quadrotor that launches from a 3-inch tube to a height of 10~m or more, and then passively unfolds to a fully functional multirotor when triggered by a nichrome burn wire release mechanism. This prior work introduced the basic aerodynamic principles and structural design concepts required to sustain the g-forces associated with a ballistic launch. A prototype was fabricated and ballistically launched from a vehicle moving at speeds of 80~km/h (22~m/s). However, the multirotor was stabilized by a remote pilot after the ballistic launch phase. 


This paper advances the line of investigation started in \cite{SQUID:IROS19} and presents the design, development and testing of a full-scale \emph{SQUID} prototype.  Capable of carrying a significant sensor payload, \emph{SQUID} transitions from a folded, 6 inch-diameter (152.4~mm) launch configuration to an autonomous, fully-controllable hexacopter after launch (Fig.~\ref{fig:fernando_mother_cake}). The entire process from launch to stabilization requires no user input and demonstrates the viability of using ballistically-launched multirotors for useful missions.

We review the full-scale \emph{SQUID} design (Section~\ref{sec:Design}), focusing on key changes from the first prototype \cite{SQUID:IROS19}. Section~\ref{sec:Launch} then describes the ballistic launch phase, Section~\ref{sec:Scaling} describes scale-model testing used to validate \emph{SQUID}'s passive stabilization design, and Section~\ref{sec:Stabilization} details the autonomous stabilization procedure. Experiments summarized in Section~\ref{sec:Testing} demonstrate the passive-to-active stabilization pipeline. Conclusions are found in Section~\ref{sec:Conclusion}.

\section{Mechanical Design} \label{sec:Design} 

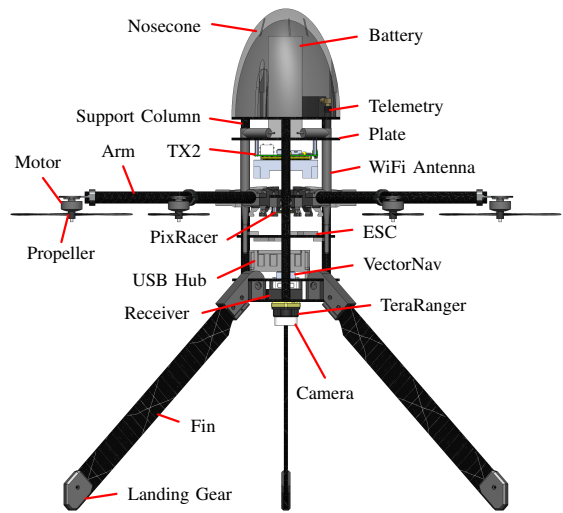
\begin{figure}[htbp]
\centering
    \begin{tikzpicture}
	    \node[anchor=south west,inner sep=0] (image) at (0,0) {\includegraphics[width=3in]{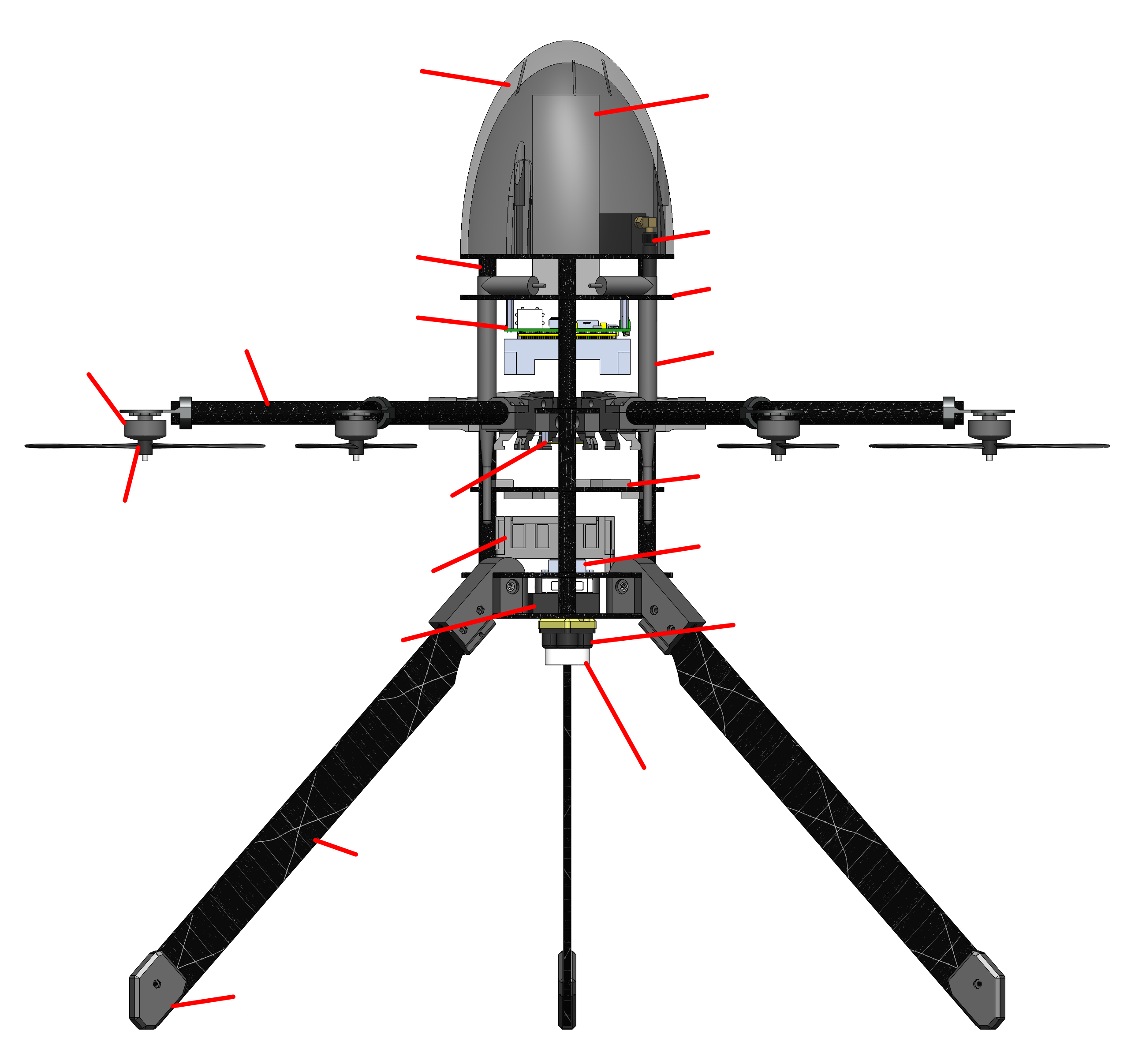}};
	    \begin{scope}[x={(image.south east)},y={(image.north west)}]
	    	\node [left ] at (0.37,0.93) {\scriptsize Nosecone};
	    	\node [right] at (0.63,0.91) {\scriptsize Battery};
	    	\node [right] at (0.63,0.78) {\scriptsize Telemetry};
	    	\node [right] at (0.63,0.73) {\scriptsize Plate};
	    	\node [right] at (0.63,0.67) {\scriptsize WiFi Antenna};
	    	\node [right] at (0.62,0.55) {\scriptsize ESC};	
	    	\node [right] at (0.62,0.49) {\scriptsize VectorNav};
	    	\node [right] at (0.65,0.41) {\scriptsize TeraRanger};	
	    	\node [below] at (0.57,0.28) {\scriptsize Camera};	
	    	\node [right] at (0.21,0.06) {\scriptsize Landing Gear};	
	    	\node [right] at (0.32,0.19) {\scriptsize Fin};	
	    	\node [left ] at (0.35,0.40) {\scriptsize Receiver};
	    	\node [left ] at (0.38,0.46) {\scriptsize USB Hub};
	    	\node [left ] at (0.40,0.54) {\scriptsize PixRacer};
	    	\node [below] at (0.11,0.54) {\scriptsize Propeller};
	    	\node [above] at (0.07,0.65) {\scriptsize Motor};
	    	\node [above] at (0.21,0.67) {\scriptsize Arm};
	    	\node [left ] at (0.37,0.70) {\scriptsize TX2};
	    	\node [left ] at (0.37,0.76) {\scriptsize Support Column};
	
	    \end{scope}
	\end{tikzpicture}
  \vspace{-0.1cm}
\caption{An annotated view of \emph{SQUID}.}
\label{fig:frame}
\end{figure}

\begin{figure}[t]
\centering
    \begin{tikzpicture}
	    \node[anchor=south west,inner sep=0] (image) at (0,0) {\includegraphics[width=2.5in]{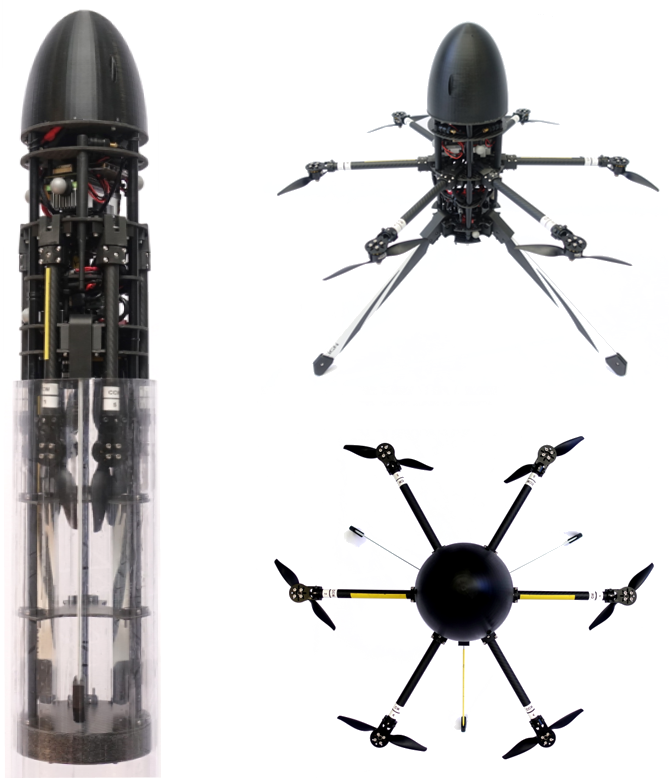}};
	    \begin{scope}[x={(image.south east)},y={(image.north west)}]
	    	\node [] at (0.13,-0.02) {\scriptsize (a)};
	    	\node [] at (0.69, 0.50) {\scriptsize (b)};
	    	\node [] at (0.69,-0.02) {\scriptsize (c)};
	    
	    \end{scope}
	\end{tikzpicture}
	\vspace{-0.1cm}
	\caption{\emph{SQUID} partially inside the launcher tube and interfacing with the carriage (a), and with its arms and fins fully deployed from a side (b) and top perspective (c).}	\label{fig:cake_on_the_beach}

\end{figure}
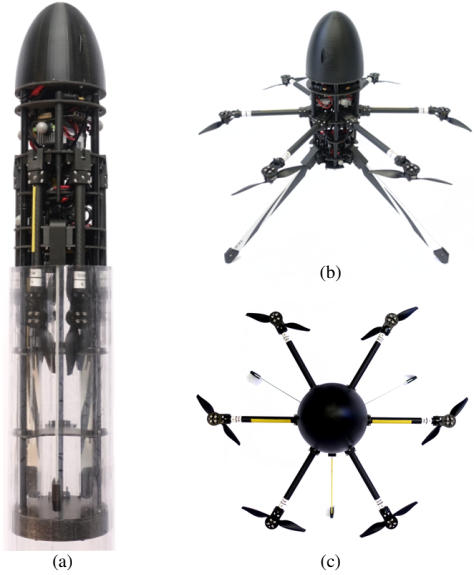

The mechanical design of the new \emph{SQUID} prototype (hereafter termed \emph{SQUID}, while $\mu$\emph{SQUID} will refer to the earlier 3-inch \emph{SQUID} prototype) is dictated by three broad functional requirements. The multirotor must: {(i)} launch from a tube (6-inch diameter for this prototype), {(ii)} travel ballistically to a predetermined height, and {(iii)} autonomously transition into stable, multirotor flight. To satisfy these non-traditional flight requirements, \emph{SQUID} blends design elements from both ballistic and multirotor platforms. The multirotor's central rigid body houses a battery and the perception and control systems, and interfaces with six fold-out arms with rotors and three fold-out fins which passively stabilize the multirotor during ballistic motion. The layout of key \emph{SQUID} components is given in Fig.~\ref{fig:frame} and the configuration in folded and deployed states are shown in Fig.~\ref{fig:cake_on_the_beach}. Table~\ref{tab:squidSummary} and Table~\ref{tab:SQUIDcomponents} provide a list of key \emph{SQUID} components and main design attributes.

\begin{table}[h!]
  \centering
  \caption{\emph{SQUID} System Properties}
  
    \begin{tabular}{lrl}
    \toprule
    Property & \multicolumn{1}{r}{Value} & \multicolumn{1}{l}{Units} \\
    \midrule
    Mass  & 3.3 &kg \\
    
    Length & 79 &cm \\
    Folded Diameter & 15 &cm  \\
    Unfolded Diameter (propeller tip-to-tip) & 58 &cm \\
    Thrust at Hover & 56&\% \\
    Launch Speed & 12 &m/s \\
    \bottomrule
    \end{tabular}%
  \label{tab:squidSummary}%
  \vspace{-0.1cm}
\end{table}%

\subsection{Central Rigid Body}
In contrast to conventional multirotors, \emph{SQUID}'s central body must sustain high transient forces during ballistic launch. Unlike prior \emph{$\mu$SQUID}, which was manually stabilized by a pilot, \emph{SQUID} also requires a perception system comprising a camera (FLIR Chameleon3), rangefinder (TeraRanger Evo 60m), IMU/barometer (VectorNav VN-100), and onboard computer (NVIDIA Jetson TX2) to achieve full autonomous stabilization. Due to these added components, the original 3D-printed aeroshell structure was abandoned in favor of a hollow carbon fiber frame in order to maximize volume, increase strength, and allow easy access to the perception and control systems.

The frame consists of six thick carbon fiber plates separated by support columns (made of aluminum standoff pins surrounded by carbon fiber tubes) that transmit the launch loads. A 3D printed nosecone reduces drag by approximately 50\% compared to a bluff body nose. The placement of the heavy LiPo battery in the nosecone shifts the center of mass (COM) upward. This placement ensures that \emph{SQUID}'s aerodynamic center (AC) trails behind the COM, which improves the passive ballistic stabilization. Passive stabilization is further addressed in Section \ref{sec:fins}.

\subsection{Rotor Arms}
The six rotors are mounted on carbon fiber tubes which attach to the central body with passive, spring-loaded hinges to allow $90^\circ$ of rotation. The arms can exist in two states: constrained by the launch tube to be parallel to the body axis (closed), or extending radially outward perpendicular to the central axis (open). For \emph{$\mu$SQUID}, the timing of the transition was controlled by an arm release mechanism \cite{SQUID:IROS19}. For \emph{SQUID} however, the transition from closed to open state occurs immediately after the multirotor leaves the launch tube, reducing mechanical complexity.

A torsional spring inside the hinge generates 1.04~N$\cdot$m of torque when the arm is closed, and half that amount when the arm is open. Vibration in the motor arms during flight dictates the addition of a spring-loaded latch to keep the arms rigidly open after deployment.

\subsection{Fins} \label{sec:fins}
\emph{SQUID}'s fins provide aerodynamic stabilization during ballistic flight to ensure the vehicle maintains the launch direction before active stabilization is engaged. Aerodynamic forces on the fins shift the multirotor's AC downward behind the COM, enabling \emph{SQUID} to passively weathercock and align with the direction of flight. Folding fins, rather than fixed fins, are a major design change from \emph{$\mu$SQUID} \cite{SQUID:IROS19} and were driven by a compromise between competing requirements of aerodynamic stability, low drag, constrained tube volume, and design simplicity. This design change was guided by the use of literature-derived expressions \cite{hoerner1958fluid, hoerner1985fluid} and scale model testing.

Fixed fins have a number of disadvantages. Any fin requires clean, unseparated flow to operate as designed. Therefore, fins that remain fixed within the tube area must also be paired with a streamlined tailbox in order to have access to said flow. This tailbox streamlining however reduces the wake drag and hence also reduces the stabilizing force it provides. Additionally, small fins which fit within the tube can only be partially effective as they have a limited wingspan. Expanding the fins along the tube only further lowers their aspect ratio (and therefore lift coefficient), reducing their capacity to move the AC. Deploying fins radially is therefore a much more effective means of enhancing stability, improving \emph{SQUID}'s ability to predictably rotate upwind.

\emph{SQUID}'s new tubular cross section and foldout fins increase stability relative to \emph{$\mu$SQUID} and simplify launch packaging issues with a simple cylindrical geometry, but do so at the cost of more ballistic drag. For most \emph{SQUID} applications however, ballistic efficiency can be sacrificed for these gains. Foldout fins can be tailored to provide a desired stability margin between the COM and AC, and provides margin for swappable payloads that may shift the COM. Given our selected 30~cm fins, the AC is located 38~cm from the nose, with a margin of 14~cm from the COM. Uncertainties in aerodynamic coefficients, drag on the arms, and the dynamics of the unfolding components can lead to substantial deviations from this calculated margin however. Accordingly, we validated our aerodynamic stability with a 3:1 scale model (50~mm diameter, 150~grams) using an open air wind tunnel (see Section \ref{sec:Scaling}) prior to full-scale tests.

While the hinges connecting the fins to the body are similar to the arm hinges, the fins do not use a latching mechanism because vertical vibrations have little impact on their functionality. ``Feet" attached to the ends of the fins protect the tips and enable them to double as landing gear.

\begin{table}[htbp]
\vspace{-0.1cm}
  \centering
  \caption{Key \emph{SQUID} components}
    \begin{tabular}{lrr}
    \toprule
    Component & \multicolumn{1}{l}{Description} & \multicolumn{1}{l}{Mass (g)}\\
    \midrule
    \textbf{\textit{Flight Electronics}}\\
    \hspace{3mm}Motors & \multicolumn{1}{l}{T-Motor F80 Pro, 1900kv} & \hspace{-100cm} 36 (x6)\\
    \hspace{3mm}ESCs & \multicolumn{1}{l}{T-Motor F30A 2-4S} & 6 (x6)\\
    \hspace{3mm}Propellers & \multicolumn{1}{l}{7" diameter x 4" pitch} & 8 (x6)\\
    \hspace{3mm}Flight Controller & \multicolumn{1}{l}{mRo PixRacer (PX4 Flight Stack)} & 11\\
    \hspace{3mm}Receiver & \multicolumn{1}{l}{X8R 8-Channel} & 17\\
    \hspace{3mm}Telemetry & \multicolumn{1}{l}{HolyBro 100 mW, 915 MHz} & 28\\
    \hspace{3mm}Battery & \multicolumn{1}{l}{4S LiPo, 6000 mAh, 50C} & 580\\
    \textbf{\textit{Perception System}}\\
    \hspace{3mm}Onboard Computer & \multicolumn{1}{l}{NVIDIA TX2} & 144\\
    \hspace{3mm}Carrier Board & \multicolumn{1}{l}{Orbitty Carrier Board}  & 41\\
    \hspace{3mm}Rangefinder & \multicolumn{1}{l}{TeraRanger Evo 60mm}  & 9\\
    \hspace{3mm}IMU/Barometer & \multicolumn{1}{l}{VectorNav VN-100}  & 4\\
    \hspace{3mm}Camera & \multicolumn{1}{l}{FLIR Chameleon3 w/ 3.5~mm Lens}  & 128\\
    \bottomrule
    \end{tabular}%
  \label{tab:SQUIDcomponents}%
  \vspace{-0.15cm}
\end{table}%

\section{Ballistic Launch Process and the Autonomous Transition to Stabilized Flight} \label{sec:Launch}

\begin{figure}[htbp]
\centering

	\begin{tikzpicture}
	    \node[anchor=south west,inner sep=0] (image) at (0,0) {\includegraphics[width=0.65\columnwidth]{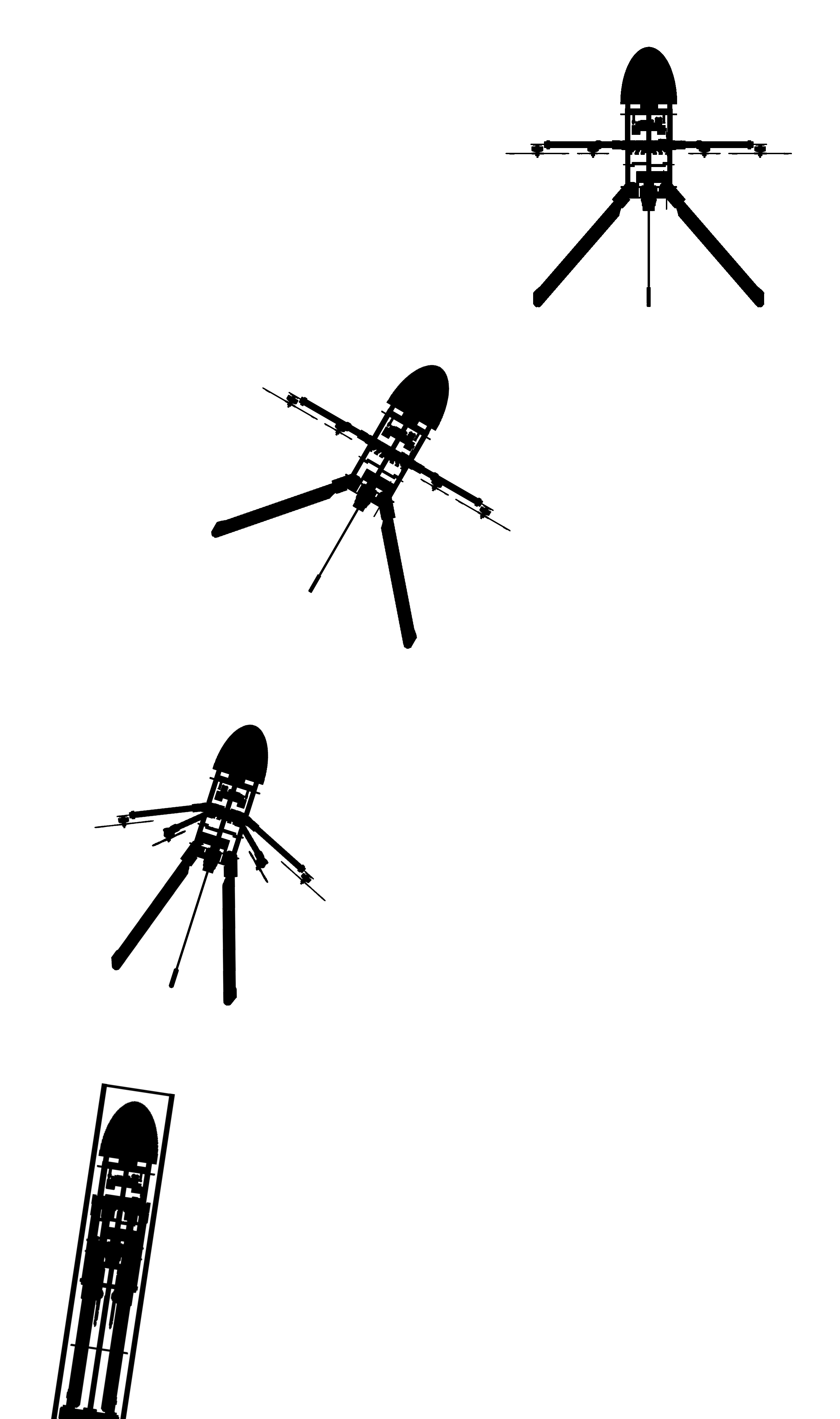}};
	    \begin{scope}[x={(image.south east)},y={(image.north west)}]
	    
	    	\node [font=\scriptsize,right,align=left ] at (0.20,0.10) {Folded \\ Configuration};
	    	\node [font=\scriptsize]                   at (0.08,0.27) {\textit{Launch}};
	    	\node [font=\scriptsize,right ,align=left] at (0.35,0.33) {Arms and Fins \\ Deploy};
	   		\node [font=\scriptsize,align=center]      at (0.25,0.55) {\textit{Ballistic} \\ \textit{Flight}};
	    	\node [font=\scriptsize,align=center]      at (0.60,0.59) {Motors \\ Activate};
	    	\node [font=\scriptsize,align=center]      at (0.44,0.80) {\textit{Active} \\ \textit{Stabilization}};	
	    	\node [font=\scriptsize,align=center]      at (0.77,0.76) {Controlled \\ Flight};
	    		    	
	        \draw [line width=0.6mm,color=lightgray,-{Latex[length=3mm]}] (0.17,0.24) -- (0.20,0.30);	
	        \draw [line width=0.6mm,color=lightgray,-{Latex[length=3mm]}] (0.31,0.50) -- (0.37,0.58);	
	        \draw [line width=0.6mm,color=lightgray,-{Latex[length=3mm]}] (0.53,0.75) -- (0.63,0.82);
	        	
	    \end{scope}
	\end{tikzpicture}	

\vspace{-0.1cm}
\caption{\emph{SQUID} deployment sequence.}
\label{fig:sequence}
\end{figure}
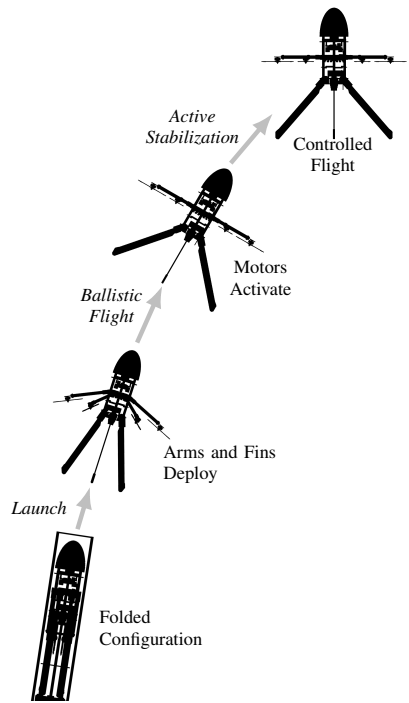

\emph{SQUID}'s mechanical design and onboard active controls manage the deployment sequence (Fig. \ref{fig:sequence}). The deployment pipeline comprises two primary phases: passive stabilization and active stabilization. In the first phase, the multirotor's aerodynamic design ensures attitude stability as it travels along a ballistic trajectory after launch. Active stabilization begins once the arms are fully deployed and occurs before the trajectory's apogee. The following sections provide details on the launch stabilization process and our experimental validation of these concepts.
 
\subsection{Ballistic Launch Process}
\emph{SQUID} is ballistically launched to a minimum height that depends on both the safety requirements of the assets near the launch site and the altitude required for the targeted investigation. All the energy needed to loft the multirotor to the desired height, as well as to overcome the drag of the passive stabilization process, must be generated over the launching tube's very short length.  Consequently, the airframe experiences very large acceleration forces while being launched. 

The core of the launch mechanism is a re-purposed T-shirt cannon~\cite{tshirtguns:online}. Pressure is supplied by a liquid CO$_2$ canister that is regulated between 5.5~bar (indoor, to stay within ceiling clearance) and 6.9~bar (outdoor, maximum safe) chamber pressure in gas phase. An aluminum stand holds the launch tube in place and allows adjustment of the launch angle. Accordingly, both the launch height and angle can be adjusted to avoid local hazards. 

Prior to launch, \emph{SQUID} rests in a folded state inside the launch tube, which is generally pointed upwards. A 300~gram carriage assembly sits between \emph{SQUID} and the tube base, transmitting launch loads generated by the compressed gas directly to the frame's support columns. A 25~mm-thick polyethylene foam disk at the base of the carriage creates a low-friction seal which maximizes the transfer of energy from the compressed gas into kinetic energy and also prevents the carriage from leaving the tube during launch. 

This launching mechanism meets requirements, but has a number of inefficiencies. After launch is triggered, the compressed gas accelerates \emph{SQUID} through the tube at approximately 21~g's (estimated from video as the IMU saturates at 16~g's), but short of the unlimited valve throughput prediction of $\approx$350~g's. The maximum height achieved with this system is also 32~m (or 1~kJ potential energy), less than a third of the imparted energy as calculated from the ideal adiabatic expansion of the CO$_2$ chamber. Discrepancies between the predicted and estimated values are thought to be from friction within the tube, a valve throughput, and air drag.

\subsection{Passive Stabilization - Launch without Wind}

After exiting the launch tube, the arms and fins deploy immediately due to the spring-loaded hinges. This deployment has four effects on the aerodynamic stability: the COM is shifted towards the nose, the AC is shifted rearward due to the fin lift, the fins increase aerodynamic damping in yaw, and mass moves outwards which increases yaw inertia.

As described in Section~\ref{sec:fins}, the lower AC helps \emph{SQUID} maintain orientation and follow the intended flight path until active stabilization begins. The large displacement between the COM and AC, coupled with the launch momentum, causes \emph{SQUID} to orient robustly into the apparent wind. When the launch tube is stationary and roughly vertical, this effect helps \emph{SQUID} to passively maintain orientation during the ballistic phase, which simplifies the transition to active stabilization.  

\subsection{Passive Stabilization - Launch in Crosswind} \label{sec:Scaling}

During launch from a moving vehicle, \emph{SQUID} experiences a strong crosswind, and will weathercock its nose in the direction of the launch platform's motion. Accordingly, \emph{SQUID}'s passive stabilization design ensures that the multirotor travels smoothly during the ballistic phase and that its orientation at the beginning of the active stabilization phase is predictable.

To validate \emph{SQUID}'s expected passive aerodynamic behavior before field testing, sub-scale wind tunnel tests were performed at the Center for Autonomous Systems and Technologies (CAST) at Caltech. These tests were intended to prove that the new folding fin architecture could provide a sufficient stabilizing effect in the presence of a crosswind.

\begin{figure}[htbp]
\centerline{\includegraphics[width=1.5in]{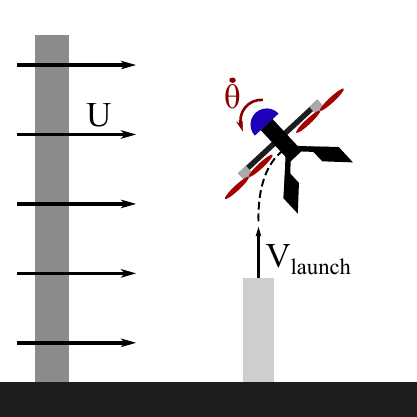} \includegraphics[width=1.5in]{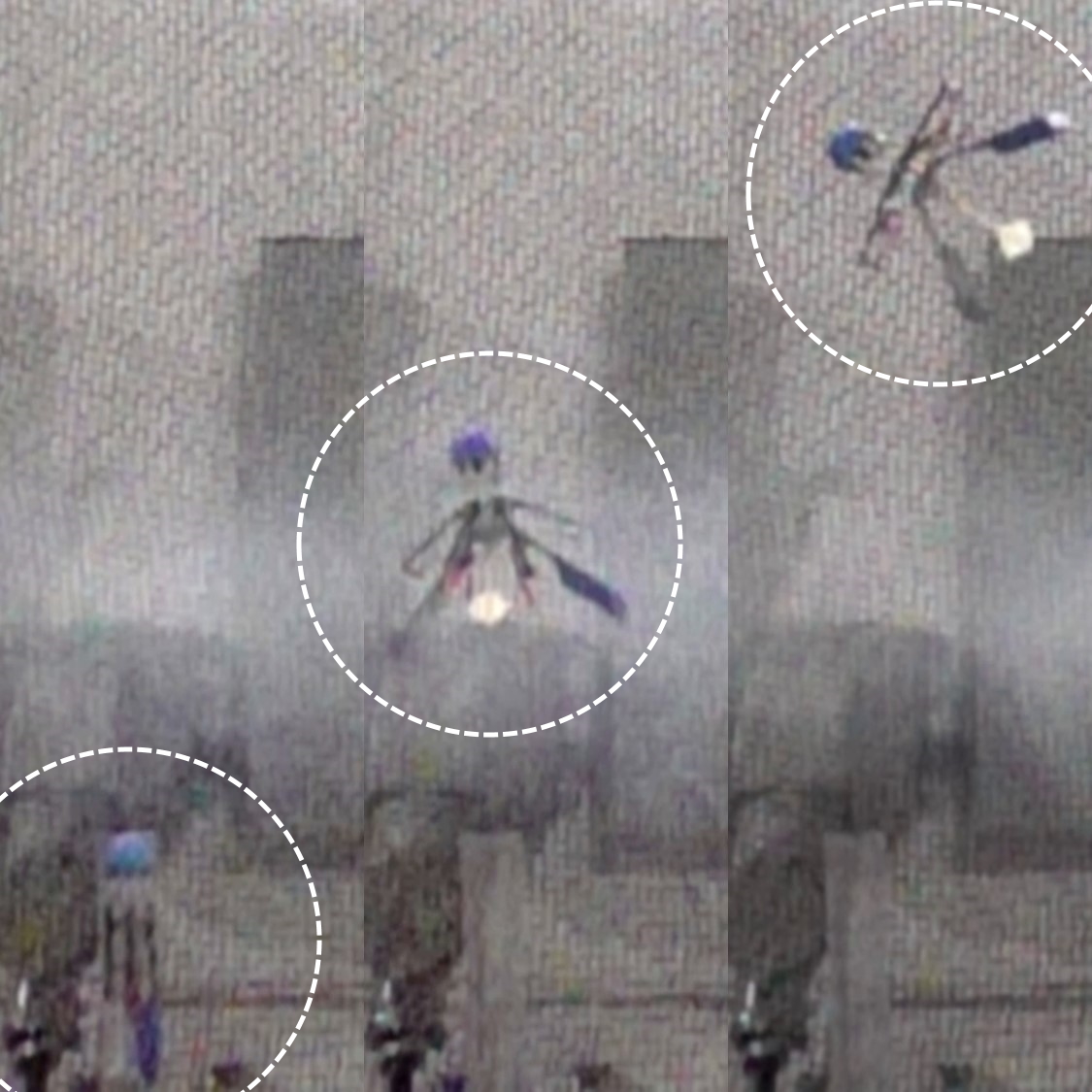}  \vspace{-0.1cm}}
\caption{Wind Tunnel Testing. Left - Definition of experiment parameters. Right - Snapshot sequence showing stable upwind pitching of the 1/3 \emph{SQUID} model.}
\label{fig:windtunnel}
\end{figure}

The sub-scale wind tunnel tests were performed using a 1/3 scale model of \emph{SQUID}. Scaling for ballistically-launched drones near apogee, discussed in greater detail in \cite{SQUID:IROS19}, primarily depends upon the Froude number ($U/\sqrt{gL}$), launch- to wind-velocity ratio, geometric parameters, and launch angle. Since \emph{SQUID}'s tailbox is a bluff-body disc, separation at the base is virtually guaranteed, meaning Reynolds effects can be neglected \cite{hoerner1958fluid}. To correct the sub-scale results to be representative of the full-scale model, the trajectories and velocities were scaled by a factor of 3 and $\sqrt{3}$, respectively~\cite{SQUID:IROS19}.

Accordingly, the performance of a vertical launch of 4.5~m/s in 10~m/s crosswinds (Fig. \ref{fig:windtunnel}) can be extrapolated to the behavior of a full-sized drone launched at 7.8~m/s in a 17~m/s crosswind. The aerodynamically stable behavior, as indicated by the upwind turn, illustrates that the multirotor with deployed fins and motor arms produces a sufficient righting moment to predictably orient the multirotor upwind on launch. While not perfectly analogous (full-scale tests were performed at 12~m/s and a slightly different geometry), these sub-scale trajectories had a similar one-third scaled stability margin (5cm) and provided confidence that the full-sized \emph{SQUID} would have a predictable trajectory if launched from a moving vehicle (a goal for future work).

\subsection{Transition from Passive to Active Stabilization}

\emph{SQUID} commences the autonomy pipeline once the distance sensor indicates the vehicle has cleared the launch tube. The passive-to-active transition occurs after the vehicle has exited the tube and the arms are fully deployed, allowing the motors to spin. Starting the motors early in the ballistic phase of launch is important as the motors need to be fully spooled up and stabilizing the multirotor before apogee. At apogee, the airspeed may not be sufficient to provide enough aerodynamic stabilization, risking the multirotor entering a tumbling state from which is may not recover.  

\section{Active Stabilization}\label{sec:Stabilization}
Our active stabilization solution is based upon previous research into autonomously recovering a monocular vision-based quadrotor after its state-estimator fails due to a loss of visual tracking~\cite{scaramuzzaReinitialization,brockersTowards}. For our visual inertial odometry pipeline, we utilize the open-source Robust Visual Inertial Odometry (ROVIO), an extended Kalman Filter that tracks both 3D landmarks and image patch features~\cite{rovio}. Since it tightly integrates image intensity information with intertial data to produce odometry estimates, ROVIO is capable of operating in stark, low-texture environments such as over pavement, water, and the surface of other planets.

The first stage of the active stabilization phase controls the attitude to a nominal zero-roll/pitch orientation using the IMU-based attitude estimate. As the air pressure around the multirotor spikes on launch, the barometric altitude estimates become unreliable and the altitude must be maintained open-loop, biased upwards for safety. The barometric readings stabilize within three seconds of launch, and at this point, \emph{SQUID} begins actively controlling its altitude and attempts to reduce the vertical velocity to zero. As no horizontal position or velocity information is available, active control of the lateral position is not possible and \emph{SQUID} continues to drift in plane until the VIO can be initialized.

Several conditions need to be met before the VIO can be successfully initialized. Firstly, the pitch and roll rates need to be near-zero to ensure that the camera captures frames with low motion blur. Secondly, the vertical velocity needs to be near-zero so the distance between the multirotor and the ground remains constant and the initial feature depth can be well established using rangefinder measurements. Finally, the lateral velocity must be small (once again to minimize motion blur), so the multirotor is allowed to drift for 10~s post spool up to enable aerodynamic drag to bleed off excess speed. Future iterations of the autonomy pipeline will sense when to initialize VIO directly from the detected motion blur, enabling the vehicle to enter position stabilization sooner after launch. 

The VIO is considered initialized when the cumulative variance of the VIO's x- and y-position estimates drop below a preset threshold. The pose estimates are then fed into the flight controller state estimator filter to be fused with the IMU. At this point, \emph{SQUID} has full onboard state estimation and can now control both altitude and lateral position. 

\subsection{Experimental Validation} \label{sec:Testing}

\begin{figure}[h!]
\centering

	\begin{tikzpicture}
	    \node[anchor=south west,inner sep=0] (image) at (0,0) {\includegraphics[width=0.7\columnwidth]{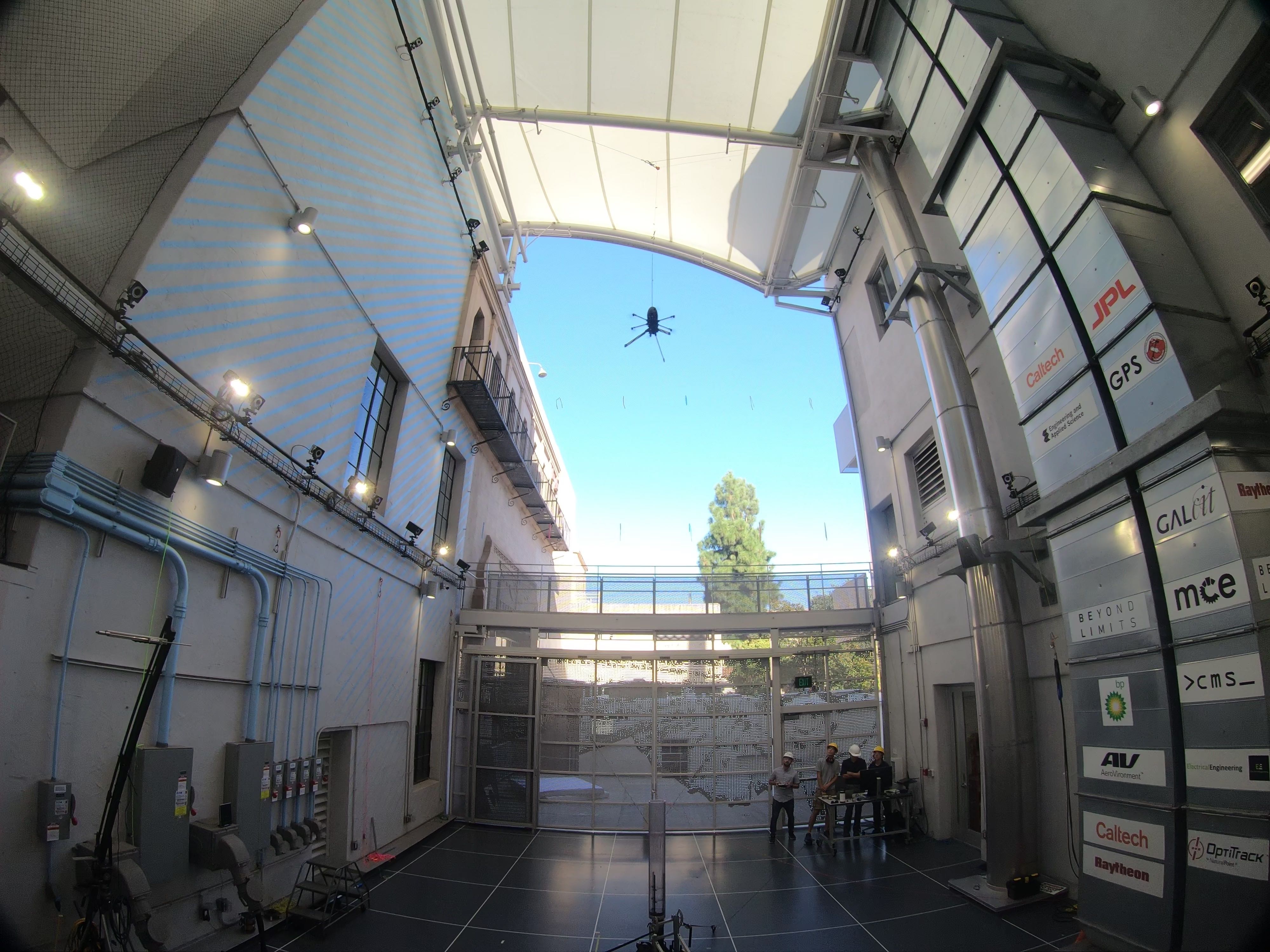}};
	    \begin{scope}[x={(image.south east)},y={(image.north west)}]
	    
	    	\node [font=\scriptsize,left,align=right] (text_Launcher) at (-0.01,0.10) {Launcher};
	    	\node [font=\scriptsize,left,align=right] (text_Cameras1) at (-0.01,0.50) {Lower Optitrack \\ Camera Rail};
	    	\node [font=\scriptsize,left,align=right] (text_Cameras2) at (-0.01,0.90) {Upper Optitrack \\ Camera Rail};
	    	\node [font=\scriptsize,left,align=right] (text_tether)   at (-0.01,0.75) {Tether Redirect};
	    	\node [font=\scriptsize,left,align=right] (text_SQUID)    at (-0.01,0.65) {\emph{SQUID}};
	    	
	        \node[circle,draw,minimum size=0.8cm,color=red,line width=0.4mm] (circ_Launcher)   at  (0.52,0.08) {};
	        \node[circle,draw,minimum size=0.5cm,color=red,line width=0.4mm] (circ_SQUID)      at  (0.515,0.65) {};
	        \node[circle,draw,minimum size=0.5cm,color=red,line width=0.4mm] (circ_Optitrack1) at  (0.20,0.57) {};
            \node[circle,draw,minimum size=0.3cm,color=red,line width=0.4mm] (circ_Optitrack2) at  (0.34,0.88) {};
            \node[circle,draw,minimum size=0.3cm,color=red,line width=0.4mm] (circ_tether) at  (0.51,0.83) {};

	    	\draw [line width=0.4mm,color=red,-{Latex[length=3mm]}] (text_Launcher) -- (circ_Launcher);
	    	\draw [line width=0.4mm,color=red,-{Latex[length=3mm]}] (text_SQUID)    -- (circ_SQUID);
	    	\draw [line width=0.4mm,color=red,-{Latex[length=3mm]}] (text_Cameras1) -- (circ_Optitrack1);
	    	\draw [line width=0.4mm,color=red,-{Latex[length=3mm]}] (text_Cameras2) -- (circ_Optitrack2);
	    	\draw [line width=0.4mm,color=red,-{Latex[length=3mm]}] (text_tether)   -- (circ_tether);
	    	
	    	
	        \node[anchor=north east,inner sep=0] (image2) at (1.0,1.0) {\includegraphics[width=0.25\columnwidth]{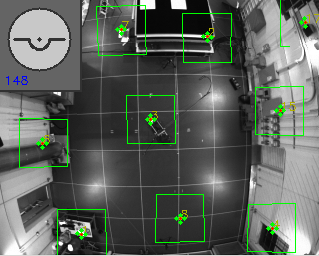}};
	        
	    	\node [font=\tiny,below left,align=right,white] at (1.0,1.0) {ROVIO View};
	    	    
	    \end{scope}
	\end{tikzpicture}	
\caption{Launching \emph{SQUID} inside CAST.} \label{fig:CAST} 
\end{figure}
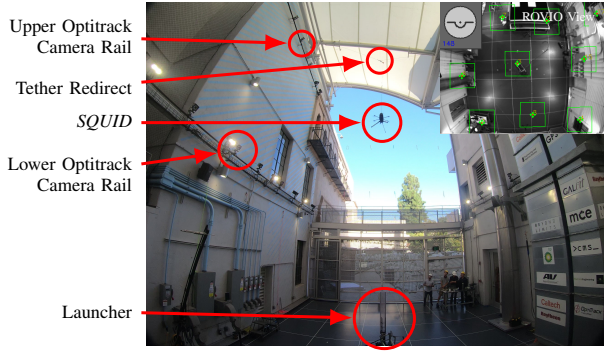

To demonstrate the proposed passive-to-active stabilization pipeline, we launched \emph{SQUID} in a 42 foot-tall flying arena at CAST (Fig.~\ref{fig:CAST}). The arena has two tiers of Optitrack motion capture cameras allowing \emph{SQUID}'s position and orientation to be tracked throughout the duration of a flight for offline analysis. During initial development, a tether system was constructed inside the arena to prevent the multirotor from damaging the facility in the event of a launch failure. A small weight was used to passively eliminate any slack in the tether. As SQUID accelerates significantly faster than the 1~g of the counterweight (note the slack in the tether in Fig.~\ref{fig:fernando_mother_cake}), it is unlikely that the tether interfered with the critical passive-to-active attitude stabilization phase.

\begin{figure}[htbp]
\centering
\includegraphics[scale=0.9]{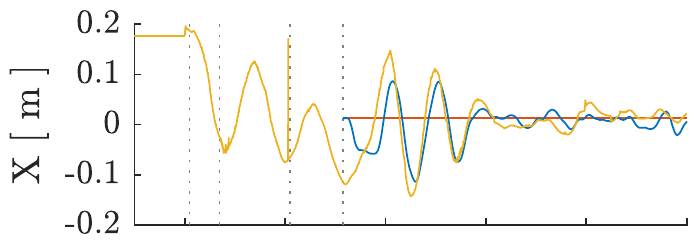} \\
\includegraphics[scale=0.9]{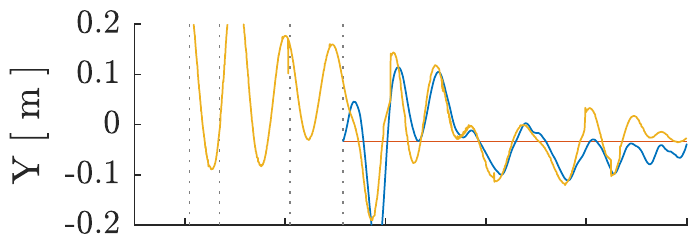} \\
\includegraphics[scale=0.9]{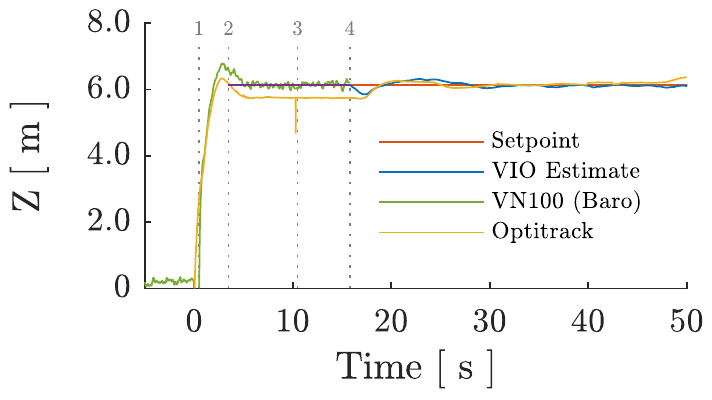} \vspace{-0.1cm}
\caption{Onboard state estimates and ground truth during launch. 1: Motors on, 2: Closed-Loop altitude control, 3: VIO initialization, 4: Position control} \label{fig:launch_pose}  \vspace{-0.5cm}
\end{figure}

Fig.~\ref{fig:launch_pose} shows the position tracking of a full launch to active position stabilization test flight. At launch (t=0), altitude is quickly gained as the multirotor accelerates. The motors turn on at Point 1 and begin actively stabilizing the attitude. By Point 2, the barometer has recovered from the launch and closed-loop altitude control commences. Ten seconds after the motors are turned on (Point 3), VIO initialization begins. At Point 4, the VIO is initialized and starts to feed pose estimates to the flight controller, which then actively controls the position of the multirotor, completing the pipeline. The pipeline was successfully demonstrated across several days, lighting conditions, and launch pressures.  Footage of the launches can be found at \url{https://youtu.be/mkotvIK8Dmo}.

\section{Conclusion} \label{sec:Conclusion}
\emph{SQUID} has successfully demonstrated the ability to ballistically launch and transition into autonomous onboard control. In particular, we demonstrate:
\begin{enumerate}
    \item A 3.3~kg hexacopter with a payload of an advanced sensor package and mission computer.
    \item An airframe strong enough to carry and transmit launch loads without damaging onboard components.
    \item Passive aerodynamic stability generated by folding fins that set the necessary preconditions for transition to autonomous flight.
    \item Wind tunnel testing that validates the proposed multirotor design in cross-wind launches.
    \item An autonomy pipeline that carries the platform from launch detection to full 6-degree of freedom stabilization using only onboard sensing (IMU, barometer, rangefinder, and camera) and without the need for GPS.
\end{enumerate}

\begin{figure}[h!]
\centering
    \includegraphics[height=1.5 in]{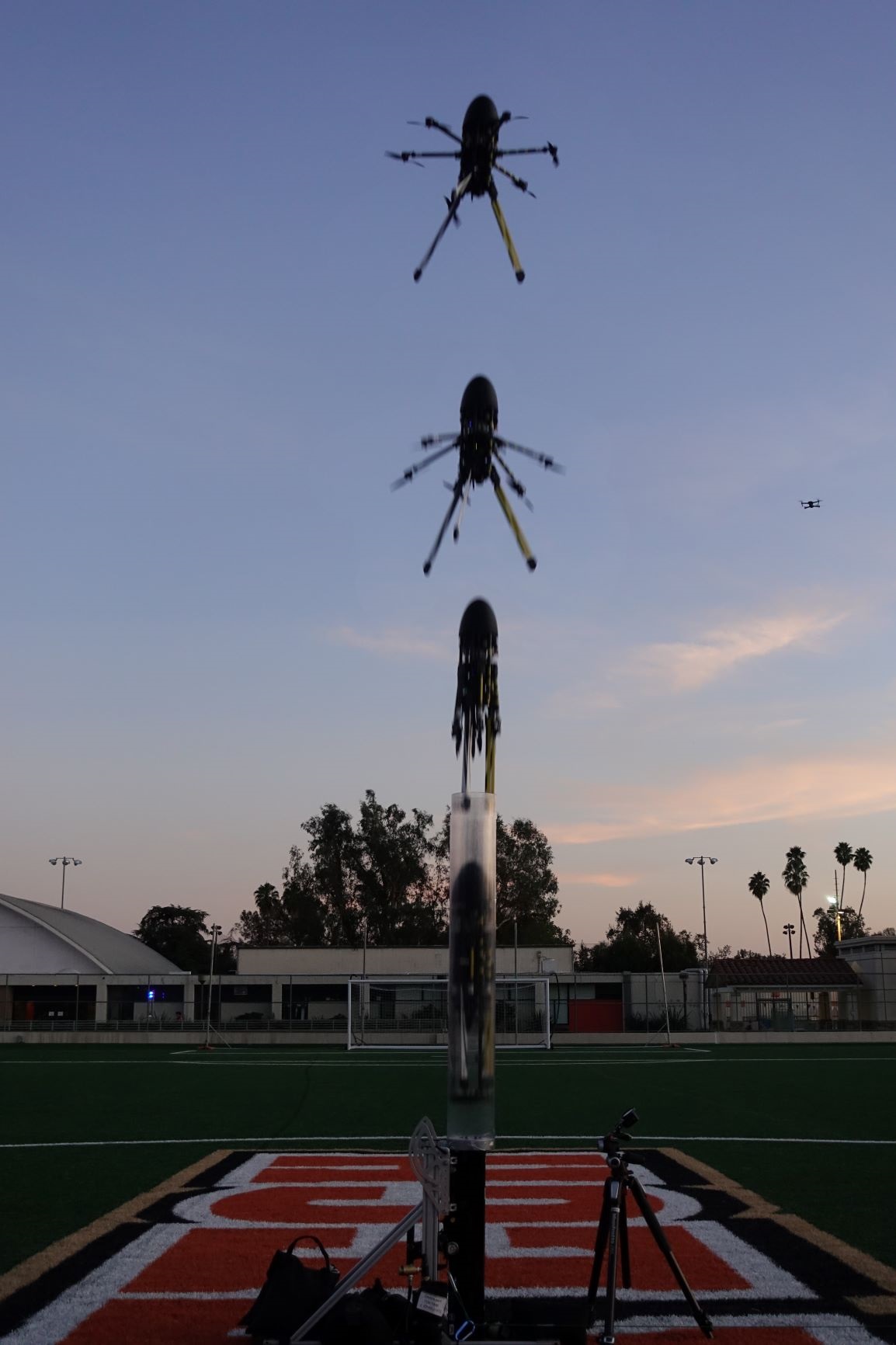}
    \begin{tikzpicture}
	    \node[anchor=south west,inner sep=0] (image) at (0,0) {\includegraphics[height=1.5 in]{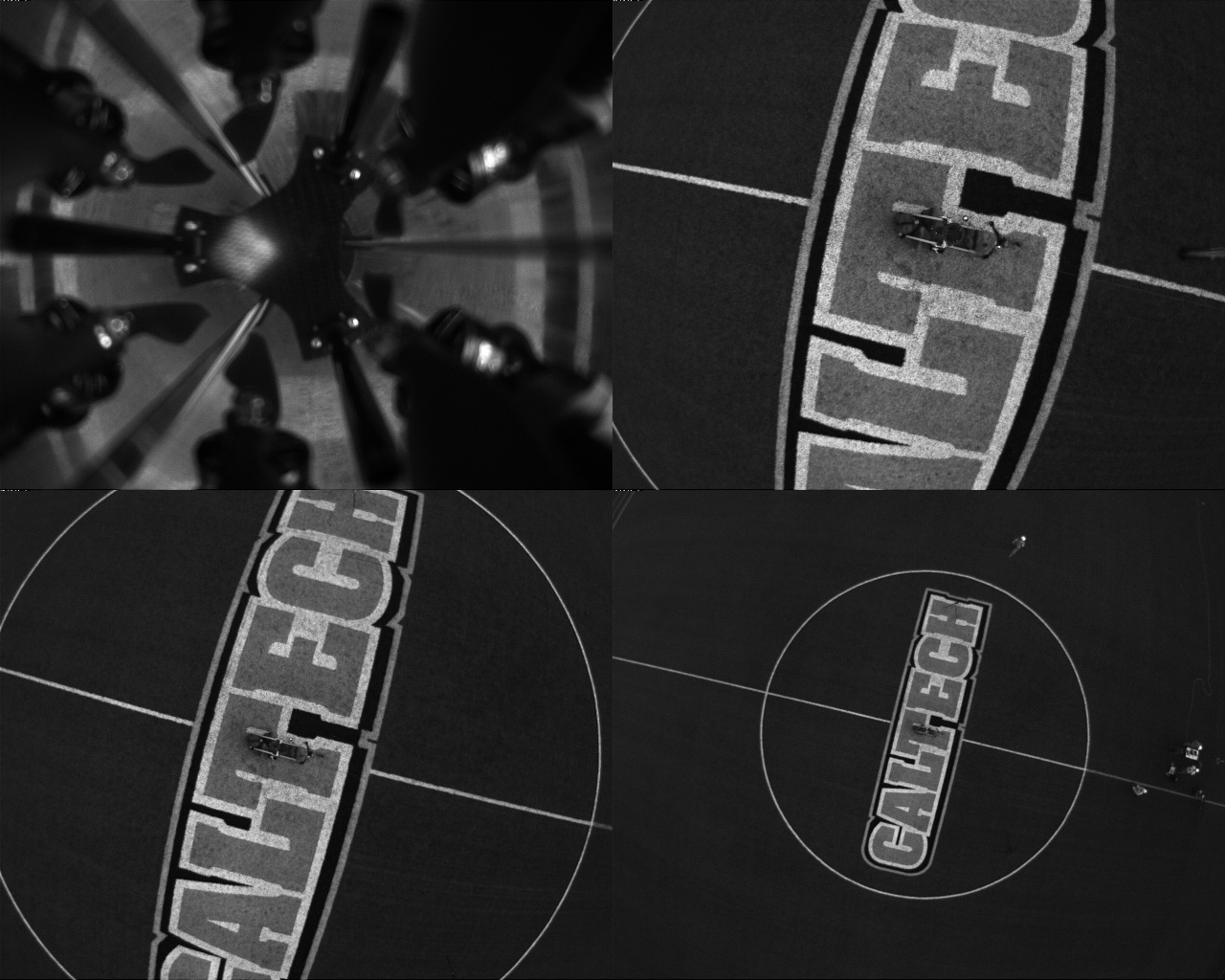}};
	    \begin{scope}[x={(image.south east)},y={(image.north west)}]
	    
	    	\node [font=\scriptsize,above left,align=right,white] at (0.5,0.5) {0.07 s};
	    	\node [font=\scriptsize,above left,align=right,white] at (1.0,0.5) {0.13 s};
	    	\node [font=\scriptsize,above left,align=right,white] at (0.5,0.0) {0.20 s};
	    	\node [font=\scriptsize,above left,align=right,white] at (1.0,0.0) {1.13 s};
	    	
	    	\draw [line width=0.2mm,white] (0.0,0.5) -- (1.0,0.5);
	    	\draw [line width=0.2mm,white] (0.5,0.0) -- (0.5,1.0);

	    \end{scope}
	\end{tikzpicture}	
\caption{Preliminary outdoor free-flight \emph{SQUID} testing.} \label{fig:outdoor_launches} 
\end{figure}
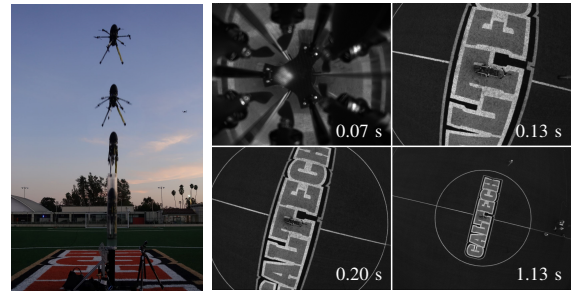

To further validate the robustness of the presented system, future development of \emph{SQUID} will include outdoor launches in windy/gusty conditions (Fig.~\ref{fig:outdoor_launches}) and launches from a moving vehicle. Planned hardware improvements include a delayed fin- and arm-release trigger to extend the ballistic range. 

This proof-of-concept system validates the viability of a ballistically-launched multirotor that deploys without human involvement, opening up new applications in fields such as disaster response, defense, and space exploration. 

\addtolength{\textheight}{-2cm}   




\section*{Acknowledgements}

This research was funded by the Jet Propulsion Laboratory, California Institute of Technology, under a contract with the National Aeronautics and Space Administration. This research was also developed with funding from the Defense Advanced Research Projects Agency (DARPA). The authors also thank Marcel Veismann, Andrew Ricci, and Robert Hewitt.

\bibliographystyle{IEEEtran}
\bibliography{references}

\end{document}